\documentclass[aps,prl,showpacs,twocolumn]{revtex4}
\usepackage{amsmath, amssymb}
\usepackage{graphicx}
\usepackage{subfigure}

\newcommand{\cici}{{\rm\raisebox{-2.pt}{$\>\stackrel{\scriptstyle\circ}{\scriptstyle\circ}\>$}}}

\begin{document}

\title{Nonclassical correlation properties of radiation fields}

\author{Werner Vogel} \email{werner.vogel@uni-rostock.de}
\affiliation{Arbeitsgruppe Quantenoptik, Institut f\"ur Physik, Universit\"at
  Rostock, D-18051 Rostock, Germany}

\begin{abstract}
  A full characterization of nonclassical space-time dependent correlations of
  radiation is formulated in terms of normally and time-ordered field
  correlation functions. It describes not only the properties of initially
  prepared multimode radiation fields, but also the dynamics of radiation
  sources.  Some of these correlation effects occur in the resonance
  fluorescence of a single two-level atom.
\end{abstract}


\maketitle

Nonclassical effects of radiation have not lost any part of their attraction
since the early days of quantum
physics~\cite{annphys17-132,natwi23-807,pr-44-777}.  It took over seven
decades until Einsteins postulate of the existence of photons could be
verified by a clear demonstration of the antibunching of
photons~\cite{prl-39-691}. Other demonstrations of nonclassical radiation
properties, such as sub-Poissonian photon statistics~\cite{prl-51-384} and
quadrature squeezing~\cite{prl-55-2409} were following soon.

In the field of quantum optics the study of nonclassical effects of radiation
was widely based on the Glauber-Sudarshan
$P$-function~\cite{prl-10-277,pr-131-2766}. Whenever it fails to have the
properties of a probability density, then the state is said to be
nonclassical~\cite{pr-140B-676,scr-T12-34}. This condition describes 
prominent examples of nonclassical effects.

Recently some possibilities of a complete characterization of nonclassicality
have been developed for a single mode of the radiation field. Observable
conditions could be derived, which are based on characteristic
functions~\cite{prl-84-1849,prl-89-283601} or on
moments~\cite{pra-71-011802,pra-72-043808}. Both approaches lead to infinite
hierarchies of nonclassicality conditions in terms of the corresponding
quantities. The characteristic function approach has already been used
in some
experiments~\cite{pra-65-033830,quant-ph/0704.0179}.

Nonclassicality of multimode radiation has been considered in some special
cases only. Entanglement, a special nonclassical property, is considered to be
useful for quantum information processing.  A complete characterization in
terms of moments is known for bipartite continuous-variable entangled states
whose partially transposed density operator exhibits some
negativities~\cite{prl-95-230502}, for some special cases see
also~\cite{njp-7-211,prl-96-050503}. There exist attempts to generalize the
method for the multimode case~\cite{pra-74-030302} and for bound
entanglement~\cite{quant-ph/0605001}.

In the present Letter we provide a general characterization of nonclassical
correlation properties of radiation fields in terms of space-time dependent
(normally and time-ordered) field correlation functions.  Beyond the
characterization of initially prepared quantum states of free multimode
radiation, this approach also describes the nonclassical effects caused by the
dynamics of radiation sources.  The physical realization and the detection of
the new correlation effects are also considered.

Let us deal with field correlation properties in an arbitrary but fixed number
of $k$ space-time points. In practice this number will depend on the used
detection setup. Consider an operator function $\hat{f}$,
\begin{equation}\label{eq:f-mult}
\hat{f} \equiv \hat{f}
[\hat{E}^{(-)}(1),\dots,\hat{E}^{(-)}(k),\hat{E}^{(+)}(k),\dots,\hat{E}^{(+)}(1)],
\end{equation}
depending on the positive and negative frequency parts of the electric field operators, $\hat{E}^{(\pm)}(i)$, where $i\equiv
({\bf r}_i,t_i)$ is the space-time argument ($i=1,\dots,k$).
We expand the operator $\hat{f}$ as
\begin{eqnarray}
\label{eq:f-taylor-mult}
\hat{f} = \sum_{\{n_i,m_i\}=0}^\infty &c_{\{n_i,m_i\}}
[\hat{E}^{(-)}(1)]^{n_1}\dots[\hat{E}^{(-)}(k)]^{n_k} \nonumber\\ &\times
[\hat{E}^{(+)}(k)]^{m_k}\dots [\hat{E}^{(+)}(1)]^{m_1},
\end{eqnarray}
where the notation ${\{n_i,m_i\}}$ is used for the dependence on
$n_1,\dots,n_k,m_1,\dots,m_k$.  

Classicality of normally and time-ordered
correlation properties can be defined as
\begin{equation}\label{eq:class-mult}
\forall  \hat{f}: \quad \langle \cici \hat{f}^\dagger \hat{f} \cici \rangle \ge 0,
\end{equation}
which generalizes the single-mode condition~\cite{pra-72-043808}.  The $\cici
\dots \cici $ notation (cf.~\cite{book}) denotes both normal ($:\dots:$
notation) and time-ordering, with increasing times in the fields
$\hat{E}^{(-)}$ and $\hat{E}^{(+)}$ from left to right and right to left,
respectively. By generalizing the standard definition of the $P$-function
(see~\cite{prl-10-277,pr-131-2766,book}) to the functional
\begin{equation}
 P[\{E^{(+)}(i)\}] =\langle \cici \prod_{i=1}^k \hat{\delta}
 (\hat{E}^{(+)}(i)-E^{(+)}(i)) \cici \rangle,
\end{equation}
the field correlation functions can be formally written as those of 
classical stochastic processes. They behave like classical ones, if
$P[\{E^{(+)}(i)\}]$ has the properties of a classical joint probability
density.  In the classical limit the operators correspond to classical
quantities: $\hat{E}^{(\pm)}(i) \to {E}^{(\pm)}(i)$ and $\hat{f} \to f$,
ordering prescriptions are no longer needed, and averages become classical
ones, $\langle \dots \rangle_{\rm cl}$. In this case the left-hand side (lhs)
of Eq.~(\ref{eq:class-mult}),
\begin{equation}\label{eq:class-mult-a}
\langle \cici \hat{f}^\dagger \hat{f} \cici \rangle \to \langle |f|^2
\rangle_{\rm cl},
\end{equation}
is indeed nonnegative in general. 
 
Based on this results, a radiation field shows nonclassical normally and
time-ordered correlation properties in $k$ space-time points, iff
\begin{equation}\label{eq:nonclass-mult}
\exists  \hat{f}: \quad  \langle \cici \hat{f}^\dagger \hat{f} \cici \rangle < 0.
\end{equation}
This completely defines the nonclassical correlation effects in the chosen
space-time points. It is difficult, however, to handle the nonclassicality
condition in this form. The operator $\hat{f}$ must be considered for all
choices of the coefficients $c_{\{n_i,m_j\}}$.  Moreover, it is unclear how to
observe the quantity $\langle \cici \hat{f}^\dagger \hat{f} \cici \rangle$ for
any operator $\hat{f}$.

We reformulate the condition~(\ref{eq:nonclass-mult}) solely in terms of
field correlation functions.  Inserting Eq.~(\ref{eq:f-taylor-mult}) into the
lhs of Eq.~(\ref{eq:nonclass-mult}), we obtain a quadratic form:
\begin{widetext}
\begin{eqnarray}
\label{eq:quadr-form-mult}
\langle \cici \hat{f}^\dagger \hat{f} \cici \rangle 
=\sum_{\{p_i,q_i,n_i,m_i\}=0}^\infty c^\ast_{\{p_i,q_i\}}c_{\{n_i,m_i\}}
\langle \cici
[\hat{E}^{(-)}(1)]^{n_1+q_1}\dots[\hat{E}^{(-)}(k)]^{n_k+q_k}
[\hat{E}^{(+)}(k)]^{m_k+p_k}\dots [\hat{E}^{(+)}(1)]^{m_1+p_1} \cici \rangle.
\end{eqnarray}
\end{widetext}
The necessary and sufficient condition for any violation of the nonnegativity
of the quadratic form requires the negativity of at least one of the principal
minors of the form~(\ref{eq:quadr-form-mult}).  In view of the lengthy
expressions we may only outline the procedure and consider some examples.

The nonclassicality conditions obtained from negativities of the
second-order minor read as
\begin{widetext}
\begin{eqnarray}
\label{eq:noncl-loworder-mult}
\left |\left \langle \cici
[\hat{E}^{(-)}(1)]^{n_1+q_1}\dots[\hat{E}^{(-)}(k)]^{n_k+q_k}
[\hat{E}^{(+)}(k)]^{m_k+p_k}\dots [\hat{E}^{(+)}(1)]^{m_1+p_1} \cici
\right \rangle \right |^2 > \nonumber\\
\left \langle \cici
[\hat{I}(1)]^{n_1+m_1}\dots[\hat{I}(k)]^{n_k+m_k} \cici \right  \rangle 
\left \langle \cici [\hat{I}(1)]^{p_1+q_1}\dots [\hat{I}(k)]^{p_k+q_k} \cici
\right \rangle, 
\end{eqnarray}
\end{widetext}
where $\hat{I}(i) = \hat{E}^{(-)}(i)\hat{E}^{(+)}(i)$ is the intensity
operator. These conditions are sufficient for the existence of nonclassical
correlations in the chosen space-time points. The necessary and sufficient
conditions require the consideration of the minors of all higher orders. 

Let us give an example for nonclassicality conditions based on third-order
minors, leading to inequalities composed of sums of products of three
correlation functions. Specifying the minors is equivalent to
start with a simplified form of the operator $\hat{f}$, such as 
\begin{equation}
\label{eq:f-taylor-ex}
\hat{f} = c_1 [\hat{E}^{(+)}(1)]^m + c_2 [\hat{E}^{(-)}(2)]^n + c_3  :[\hat{I}(3)]^p:,
\end{equation}
with redefined coefficients $c_i$. The resulting nonclassicality condition reads as
\begin{widetext}
\begin{equation}\label{eq:minor-3}
    \begin{vmatrix}
      \langle :[\hat{I} (1)]^m : \rangle \,  & \, \langle \cici[\hat{E}^{(-)}(1)]^m
      [\hat{E}^{(-)}(2)]^n \cici
      \rangle \, & \, \langle \cici [\hat{E}^{(-)}(1)]^m [\hat{I}(3)]^p \cici
      \rangle  \\
\langle \cici[\hat{E}^{(+)}(1)]^m
      [\hat{E}^{(+)}(2)]^n \cici
      \rangle 
   \, &\,  \langle :
      [\hat{I} (2)]^n:
      \rangle \, & \, \langle  \cici [\hat{I} (3)]^p [\hat{E}^{(+)}(2)]^n \cici
      \rangle\\
 \langle \cici [\hat{E}^{(+)}(1)]^m [\hat{I}(3)]^p \cici
      \rangle  \, & 
\, \langle  \cici [\hat{I} (3)]^p [\hat{E}^{(-)}(2)]^n \cici
      \rangle
\,  & \,  \langle : [\hat{I}(3)]^{2p}:
      \rangle      
    \end{vmatrix} \! < 0.
\end{equation}
\end{widetext}
More generally, the minor contains more complex correlation
functions, such as those in the condition~(\ref{eq:noncl-loworder-mult}).

We may further simplify the conditions~(\ref{eq:noncl-loworder-mult}). By
choosing the only nonvanishing powers as $n_1=m_1=p_2=q_2=1$, we get
\begin{equation}
\label{eq:ab-nonst}
\left \langle \cici \hat{I}(1) \hat{I}(2) \cici \right \rangle   
> \sqrt{\left \langle :
[\hat{I}(1)]^2 : \right  \rangle 
\left \langle : [\hat{I}(2)]^2 :
\right \rangle}. 
\end{equation}
This is the photon-antibunching condition in its general form, which can be
also derived as a violation of the Schwarz inequality, for example
cf.~\cite{book}.  The intensity correlation function on the lhs is
nonnegative, so that the absolute value can be omitted.  In the given form the
antibunching condition also applies for nonstationary
radiation~\cite{miranowicz}. The stationary condition was used
in the pioneering photon-antibunching experiment in resonance
fluorescence~\cite{prl-39-691}.

It is straightforward to formulate higher-order generalizations of the
antibunching condition~(\ref{eq:ab-nonst}). In
Eq.~(\ref{eq:noncl-loworder-mult}) we choose for the nonvanishing powers
$n_1=m_1=N-n$, $n_2=m_2=M-m$, $q_1=p_1=n$, and $q_2=p_2=m$ ($N\ge n$ and $M\ge
m$), which leads to 
\begin{widetext}
\begin{eqnarray}
\label{eq:noncl-higher}
\left \langle \cici
[\hat{I}(1)]^{N}[\hat{I}(2)]^{M} \cici
\right \rangle  >
\sqrt{\left \langle \cici
[\hat{I}(1)]^{2(N-n)}[\hat{I}(2)]^{2(M-m)} \cici \right  \rangle 
\left \langle \cici [\hat{I}(1)]^{2n} [\hat{I}(2)]^{2m} \cici
\right \rangle}. 
\end{eqnarray}
\end{widetext}
If we further specify $m=M$ and $n=0$, the condition
\begin{equation}
\label{eq:noncl-higher-ab}
\left \langle \cici
[\hat{I}(1)]^{N}[\hat{I}(2)]^{M} \cici
\right \rangle  >
\sqrt{\left \langle :
[\hat{I}(1)]^{2N} : \right  \rangle 
\left \langle : [\hat{I}(2)]^{2M} :
\right \rangle} 
\end{equation}
represents a direct higher-order generalization of the antibunching
condition~(\ref{eq:ab-nonst}).

From the nonclassicality condition~(\ref{eq:noncl-loworder-mult}) one may also
obtain conditions for the normally and time-ordered correlations of
intensity and field strength. Such correlations have been studied in the
resonance fluorescence of a single atom~\cite{prl-67-2450} and their
measurement has been analyzed~\cite{pra-51-4160,prl-85-1855}. A recently
proposed method of balanced homodyne correlation measurements combines the
advantages of balanced homodyning with those of correlation
techniques~\cite{prl-96-200403}. By using a strong local oscillator it allows
one to detect correlation functions of higher orders, even when the overall
quantum efficiency is small.

Let us consider a typical example. Choosing in
Eq.~(\ref{eq:noncl-loworder-mult}) the non-zero powers as
$n_i =m_i+p_i$ with $p_i\ge 0$, we obtain the condition
\begin{eqnarray}
\label{eq:noncl-field-int}
\left |\left \langle \cici
[\hat{E}^{(-)}(1)]^{p_1}\dots[\hat{E}^{(-)}(k)]^{p_k}
[\hat{I}(k)]^{m_k} \dots [\hat{I}(1)]^{m_1} \cici
\right \rangle \right | \nonumber\\
 > \sqrt{\left \langle \cici
[\hat{I}(1)]^{2 m_1 +p_1}\dots [\hat{I}(k)]^{2 m_k+p_k} \cici \right
\rangle}.\quad \quad
\end{eqnarray}
It gives some insight in the nonclassical correlation properties of intensity
and field strength operators, by comparing normally and time-ordered
correlation functions containing the field strength (here the negative frequency
part) with intensity correlation functions.
In the lowest order we may choose
$p_1=m_2=1$ as the only nonvanishing powers, which yields
\begin{equation}
\label{eq:field-int}
\left |\left \langle \cici
\hat{E}^{(-)}(1)\hat{I}(2)\cici
\right \rangle \right | > 
\sqrt{\left \langle \cici
\hat{I}(1)[\hat{I}(2)]^2 \cici \right \rangle}.
\end{equation}
We can also formulate
conditions for intensity-field strength correlations of other
types. Let us restrict the
generality of the operator $\hat{f}$, for example by setting
\begin{equation}
\label{eq:f-example}
\hat{f} = c_1
\hat{E}^{(+)}(1) + c_2 \hat{I}(2). 
\end{equation}
Inserting this expression into the condition~(\ref{eq:nonclass-mult}), we
get
\begin{equation}
\label{eq:field-int-a}
\left |\left \langle \cici
\hat{E}^{(-)}(1)\hat{I}(2)\cici
\right \rangle \right | > 
\sqrt{\left \langle 
\hat{I}(1) \right \rangle\left \langle :  [\hat{I}(2)]^2 : \right \rangle}.
\end{equation}
In general it depends on the chosen radiation sources, in particular on
their intensity correlation properties, whether this form of the condition is
stronger than the form~(\ref{eq:field-int}) or vice versa.

It is also of interest to formulate nonclassicality conditions including
the full field strength.  Equating the coefficients of $\hat{E}^{(-)}(1)$ and
$\hat{E}^{(+)}(1)$ in the operator $\hat{f}$, the positive frequency part in
Eq.~(\ref{eq:f-example}) is replaced with the field strength operator,
$\hat{E}(1) =\hat{E}^{(-)}(1) +\hat{E}^{(+)}(1)$. Now the nonclassicality
condition is of the form
\begin{equation}
\label{eq:field-int-b}
\left |\left \langle \cici
\hat{E}(1)\hat{I}(2)\cici
\right \rangle \right | > 
\sqrt{\left \langle :
[\hat{E}(1)]^2 : \right \rangle\left \langle :  [\hat{I}(2)]^2 : \right
\rangle}. 
\end{equation}
The field strength-intensity correlations are no longer compared with the
quantum statistical properties of the intensity alone.  Note that nonclassical
correlations of intensity and field strength of lowest order have been
discussed under special conditions, such as for Gaussian
fluctuations~\cite{prl-85-1855}.

In the following we deal with the question of whether or not nonclassical
correlation properties of the types introduced above may occur in the
irradiation of realistic sources. As a simple example we consider the
resonance fluorescence of a single two-level atom, for the theory cf.
e.g.~\cite{book}.  In this case we have detailed knowledge of the high-order
correlation functions to be considered. In particular, the occurrence of
photon antibunching is well known. The general concept of nonclassical
correlations, however, applies to arbitrary radiation sources.  For example,
it could serve for a more complete characterization of the radiation in the
mentioned cavity experiment~\cite{prl-85-1855}, but also for many other
radiation sources.

Let us start to consider nonclassical effects based on intensity correlation
measurements of higher order, see the condition~(\ref{eq:noncl-higher}). When
at least one of the values of $N$ or $M$ becomes two or larger, both sides of
the inequality become zero, due to the fact that a single atom cannot
simultaneously emit two photons.  Higher-order nonclassical effects of this
type do not occur in resonance fluorescence from a two-level atom, their
observation requires other types of radiation sources.

Consider now the situation for field strength-intensity correlations.
Whenever a term $[\hat{E}^{(\pm)}(i)]^{n}$ with $n \ge 2$ occurs
($i=1,\dots,k$) on the lhs of the condition~(\ref{eq:noncl-loworder-mult}),
the corresponding correlation function vanishes for the single-atom resonance
fluorescence. That is, such types of nonclassical correlations do not exist.  
Based on this knowledge, however, we may formulate a variety of nonclassicality
conditions in terms of those higher-order correlation functions which do not
contain terms of the mentioned type.  This leads to
inequalities of the form
\begin{eqnarray}
\label{eq:noncl-loworder-mult-example}
\left |\left \langle \cici
\hat{E}^{(-)}(1) \dots\hat{E}^{(-)}(l)\hat{I}(l+1) \dots \hat{I}(k) \cici
\right \rangle \right | > \nonumber\\
\sqrt{\left \langle \cici
\hat{I}(1)\dots\hat{I}(l)[\hat{I}(l+1)]^2 \dots [\hat{I}(k)]^2 \cici \right  \rangle},
\end{eqnarray}
for $ 1 < l < k$. To provide a nontrivial characterization of nonclassical
correlations in single-atom resonance fluorescence, the retarded times
$t_i-r_i/c$ must not become equal for two space-time points. It is easy to
verify by resonance fluorescence theory that such nonclassical correlations
exist for a two-level atom, since the rhs of the
condition~(\ref{eq:noncl-loworder-mult-example}) becomes zero. For getting
more insight in the full time-dependent correlation properties of higher
orders on the lhs, the correlation functions of the resonance fluorescence can
be calculated by standard methods, such as the quantum regression theorem, see
e.g.~\cite{book}. An experimental demonstration clearly requires the detection
of the correlation functions occuring in the
condition~(\ref{eq:noncl-loworder-mult-example}). Note that one may consider a
manyfold of other conditions for nonclassical space-time dependent
correlations by using higher-order minors of the general quadratic
form~(\ref{eq:quadr-form-mult}), for example cf. inequality~(\ref{eq:minor-3}).

An open problem is the application of space-time dependent nonclassicality
conditions for characterizing entanglement of multimode radiation fields.
Since the $P$-functions of entangled states fail to be probability densities,
they are included in the conditions derived above. To distinguish entanglement
from other nonclassical effects requires to handle time-dependent field
commutation rules, which is a nontrivial task that requires further research.
For details on normal- and time-ordering and time-dependent commutation rules
of source-attributed quantized radiation we refer to~\cite{book}.

In summary, we have studied the physical properties of nonclassical space-time
dependent correlations: their complete characterization, the realization of
such effects in resonance flourescence, and their observation by balanced
homodyne correlation measurements. In general nonclassicality is characterized
by negativities of properly arranged minors of various orders, whose entries
are normally and time-ordered field correlation functions of arbitrarily high
orders. The conditions completely describe the nonclassical correlation
properties, including entanglement, of any realistic radiation field whose
generalized Glauber-Sudarshan $P$-functional fails to be a joint probability
density.  This concept applies not only to multimode quantum states in a
nonclassical initial preparation, but also to nonclassical properties emerging
from the dynamics of radiation sources.

The author gratefully acknowledges valuable comments by C. Di Fidio, M.
Fleischhauer, Th. Richter and E. Shchukin. This work was supported
by the Deutsche Forschungsgemeinschaft.

\end{document}